# THE EFFECT OF HIGH NITROGEN PRESSURES ON THE HABITABLE ZONE AND AN APPRAISAL OF GREENHOUSE STATES

Ramses M. Ramirez[1,2]
[1]Earth-Life Science Institute, Tokyo Institute of Technology, Tokyo, Japan
[2]Space Science Institute, Boulder, Co, USA
Corresponding author: Ramses Ramirez (rramirez@elsi.jp)

**ABSTRACT**
The habitable zone (HZ) is the main tool that mission architectures utilize to select potentially habitable planets for follow-up spectroscopic observation. Given its importance, the precise size and location of the HZ remains a hot topic, as many studies, using a hierarchy of models, have assessed various factors including: atmospheric composition, time, and planetary mass. However, little work has assessed how the habitable zone changes with variations in background nitrogen pressure, which is directly connected to the habitability and life-bearing potential of planets. Here, I use an advanced energy balance model with clouds to show that our solar system's HZ is ~0.9 – 1.7 AU, assuming a 5-bar nitrogen background pressure and a maximum 100% cloud cover at the inner edge. This width is ~20% wider than the conservative HZ estimate. Similar extensions are calculated for A – M stars. I also show that cooling clouds/hazes and high background pressures can decrease the runaway greenhouse threshold temperature to ~300 K (or less) for planets orbiting any star type. This is because the associated increase in planetary albedo enables stable climates closer to the star, where rapid destabilization can be triggered from a lower mean surface temperature. Enhanced longwave emission for planets with very high stratospheric temperatures also permits stable climates at smaller orbital distances. The model predicts a runaway greenhouse above ∼330 K for planets orbiting the Sun, which is consistent with previous work. However, moist greenhouses only occur for planets orbiting A-stars.

# 1. INTRODUCTION

Over the years, many studies have continually revisited the limits of the habitable zone (HZ)(e.g. Dole 1964; Kasting et al. 1993; Kopparapu et al. 2013; Leconte et al. 2013; Yang et al. 2013; Ramirez and Kaltenegger 2014; Wolf and Toon 2014; Wolf and Toon 2015; Haqq-Misra et al. 2016; Ramirez and Kaltenegger 2017; Bin et al. 2018; Ramirez and Kaltenegger 2018), which is the circular region around a star (or multiple stars) where standing bodies of liquid water could exist on the surface of a rocky planet (Ramirez 2018a). Properly defining the habitable zone is and will likely remain (for the foreseeable future) an important scientific consideration because it is the main navigational tool that mission architectures use to find potentially habitable planets around other stars. In particular, better estimates of the location of the inner and outer edges of the habitable zone will improve target lists for follow-up atmospheric characterization.

The outer edge of the classical $CO_2$-$H_2O$ main-sequence HZ is defined by the distance beyond which the combined effects of Rayleigh scattering and $CO_2$ condensation outweigh the maximum greenhouse effect of $CO_2$. In contrast, 1-D models predict that once a planet is close enough to the star, and mean surface temperatures rise above ~340 K, the stratosphere becomes wet enough for efficient photolysis to remove an Earth-like surface water inventory within ~ 4.5 billion years (i.e., a moist greenhouse)(Kasting et al. 1993). At even closer distances to the star, a full runaway greenhouse with even faster water loss rates can be triggered (Kasting et al. 1993).

Both 1-D and 3-D (GCM) models are instrumental tools in the calculation of HZ boundaries. The outer edge of the present day classical HZ is ~1.67 AU in our solar system whereas initial 1-D estimates of our solar system's inner edge was ~0.95 AU (Kasting et al. 1993) until models with improved estimates of water vapour absorption recalculated this value to be ~ 1 AU (Kopparapu et al. 2013; Kopparapu et al. 2014). HZ boundaries were also calculated for F – M stars (Kasting et al. 1993; Kopparapu et al. 2013), and more recently, A-stars (Ramirez and Kaltenegger 2016; Ramirez and Kaltenegger 2018).

However, a value of 1 AU for the solar system inner edge is problematic because it would have triggered a runaway (or moist) greenhouse on Earth and rendered our planet uninhabitable. In reality, these preliminary 1-D calculations assumed fully-saturated atmospheres and underestimated the negative feedback from clouds (Kopparapu et al. 2013). Subsequently, more complex 3-D models with clouds have revisited this inner edge calculation. Leconte et al. (2013) used the Laboratoire de Météorologie Dynamique (LMD) GCM to find that clouds and sub-saturated relative humidity push the solar system's inner edge to ~0.95 AU, although their model bypasses the moist greenhouse and transitions directly to the full runaway instead. In contrast, Wolf and Toon (2015) used the Community Atmosphere Model (CAM) GCM to calculate a solar system inner edge at ~0.93 AU upon triggering the moist greenhouse. Other CAM GCM calculations find the moist greenhouse for K – M stars hotter than ~ 3000 K or for all M-stars respectively (Fujii et al. 2017; Kopparapu et al. 2017; Bin et al. 2018). Thus, the disagreement between models is large and it remains unclear as to whether the moist greenhouse is achievable or not (and if it is, under which circumstances?).

To this day, most classical HZ calculations assume a 1 bar nitrogen background atmosphere (with some $CO_2$), which can be used as a proxy for present Earth's atmospheric composition (Kasting et al. 1993). However, there is no reason to limit HZ planets to 1 bar of $N_2$. In our own solar system, Venus is an Earth-sized planet that has ~4 times the atmospheric nitrogen pressure that Earth does. Moreover, the background nitrogen pressure is not only important for maintaining habitability on a planet (Goldblatt et al. 2009; Wordsworth and Pierrehumbert 2014), but atmospheric nitrogen pressure may reflect its speciation in the mantle and can directly influence the amount of oxygen released into the atmosphere (Mikhail and Sverjensky 2014). Also, the amount of atmospheric nitrogen helps determine the susceptibility of a planet to a moist or runaway greenhouse (Goldblatt et al. 2013).

We have previously evaluated the dependence of planetary mass on habitable zone limits, which attempts to scale atmospheric pressure (Kopparapu et al. 2014). However, as we mentioned in that study, a robust theory which scales planetary mass to habitable zone limits does not exist as it depends on stochastic factors during accretion, in addition to atmospheric scaling concerns, which may or may not be translatable to generalized HZ limits. In this study, I ignore such mass and volatile scaling complications and use an advanced latitudinally-dependent energy balance model with clouds to calculate new classical $CO_2$-$H_2O$ HZ limits for different background nitrogen pressures (up to 5 bar) on Earth mass planets. I compare these against equivalent results from my cloud-free 1D radiative-convective climate model. I then discuss implications for the moist greenhouse limit and compare my results to previous 1-D and 3-D studies.

## 2. METHODS

*The single column radiative-convective climate model*

The single-column radiative-convective climate model has 55 thermal infrared and 38 solar wavelengths (Ramirez and Kaltenegger 2017; Ramirez and Kaltenegger 2018). Atmospheres are subdivided into 100 vertical logarithmically-spaced layers that reach $\sim 1 \times 10^{-5}$ bar at the top of the atmosphere. When the atmosphere is warm enough for water vapour to convect, it follows a moist adiabat and relaxes to it if the moist adiabatic lapse rate is exceeded. Likewise, when atmospheres are cold enough for $CO_2$ to condense, lapse rates adjust down to the $CO_2$ adiabatic value (Kasting et al. 1993). The atmosphere is allowed to expand as temperatures warm (Ramirez et al. 2014b; Ramirez et al. 2014a).

The radiative-convective climate model utilizes the HITRAN database (Rothman et al. 2013) at lower temperatures (<300 K) and HITEMP(Rothman et al. 2010) at higher temperatures for water vapour and HITRAN for $CO_2$. Far wing absorption in the $CO_2$ 15-micron band utilizes the 4.3 micron region as a proxy (Perrin and Hartmann 1989). The BPS water vapour continuum is overlain between 0 and 19,000 $cm^{-1}$ (Paynter and Ramaswamy 2011). The model incorporates $CO_2$-$CO_2$ and $N_2$-$N_2$ CIA (Gruszka and Borysow 1997; Gruszka and Borysow 1998; Baranov et al. 2004; Wordsworth et al. 2010). I use a standard Thekeakara solar spectrum (Thekaekara 1973) for the Sun whereas the remaining A – M stars ($T_{EFF}$ = 2,600 – 9000 K) are modeled with Bt-Settl data (Allard et al. 2003).

All radiative-convective climate modeling simulations employ a solar zenith angle at 60 degrees (although zenith angle is not held constant for EBM calculations, see next section). Atmospheres are assumed to be fully-saturated. Inverse calculations are employed, as is standard with radiative-convective climate model HZ calculations. In these, I specify a surface temperature and calculate the solar flux required to sustain it. For the outer edge, stratospheric and surface temperatures were held at 155 K and 273 K, respectively, whereas $CO_2$ partial pressure was varied from $3.3 \times 10^{-4}$ to 34.7 bar (which is the saturation $CO_2$ partial pressure at 273 K). The effective fluxes that sustain a surface temperature of 273 K are computed for all spectral classes. At the inner edge, the $CO_2$ partial pressure was assumed to be an Earth-like 330 ppm (Kasting et al. 1993). The stratospheric temperature is 200 K, while the surface temperature is gradually incremented from 200 K to 647 K, which is the predicted runaway greenhouse threshold for a planet with an Earth-like water inventory according to some radiative-convective climate models (Kasting et al. 1993). In reality, a runaway greenhouse can be achieved at much lower temperatures once the net absorbed stellar flux exceeds the capability of the planet to emit excess outgoing radiation (Goldblatt et al. 2013; Leconte et al. 2013).

*The energy balance model*

The energy balance model (EBM) is an updated version of the non-grey latitudinally dependent (36 latitude bands, 5 degrees each) model described in Ramirez and Levi (2018). The EBM uses an explicit forward marching numerical scheme with a constant time-step (i.e. a fraction of a day). As with other similar advanced energy balance models (e.g., Williams and Kasting 1997; Batalha et al. 2016), planets in thermal equilibrium emit to space as much energy as they absorb from their stars. This is summarized by the following expression:

$$C \frac{\partial T(x,t)}{\partial t} - \frac{\partial}{\partial x} D(1-x^2) \frac{\partial T(x,t)}{\partial x} + OLR - L \frac{\partial M_{col}}{\partial t} = S(1-A) \qquad (1)$$

Where, $x$ is sine(latitude), $S$ is the incoming stellar flux, $A$ is the top of atmosphere albedo, $T$ is surface temperature, $OLR$ is the outgoing thermal infrared flux, $C$ represents the overall ocean-atmospheric heat capacity, $L$ is the latent heat flux per unit mass of $CO_2$ [$5.9 \times 10^5$ J/kg(Forget and Hourdin, F. and Talagrand 1998)], $M_{col}$ is the column mass of atmospheric $CO_2$ that condenses to or sublimates from the surface, and $D$ represents a calculated diffusion coefficient. A second order finite differencing scheme solves eqn. 1.

The EBM accesses lookup tables, which contain interpolated radiative quantities computed from the 1-D radiative-convective model (see above), including the stratospheric temperature, $T_{strat}(pCO_2, T, z)$, outgoing longwave radiation, $OLR(pCO_2, T)$, and the planetary albedo $A(pCO_2, T, z, a_s)$. Here, $pCO_2$ is the $CO_2$ partial pressure, $a_s$ is the surface albedo, and z is the zenith angle. The current EBM operates over a parameter space spanning $10^{-5}$ bar $< pCO_2 <$ 35 bar, 150 K $<$ T $<$ 390 K, $0 < z < 90°$, and $0 < a_s < 1$. The radiative model quantities are tabulated for $pN_2$ values of 1, 2, 5, and 10 bar.

The model is able to distinguish between land, ocean, ice, and clouds. As the atmosphere warms near and above the freezing point, water clouds form, with latitudinal cloud coverage (*c*) dictated by equation 2:

$$c = \min\left(0.72\log\left(F_C/F_E + 1\right), 1\right) \qquad (2)$$

Here, $F_c$ is the convective heat flux whereas $F_E$ is the convective heat flux for the Earth at 288 K (~90 W/m²) in our model. This equation is similar to that used in the CAM GCM (Xu, K.M. and Krueger 1991; Yang & Abbot 2014) and yields an Earth-like cloud cover of ~50% at a mean surface temperature 288 K.

As explained above, the radiative-convective climate model overestimates absorption at the inner edge because it neglects the negative feedback of clouds at higher temperatures and assumes 100% relative humidity. As we have done before (Ramirez and Levi 2018), this is addressed by increasing the outgoing longwave radiation (OLR) by a scaling factor (*addir*) that simulates these two effects, cooling the planet. For the Earth, incrementing OLR by 6.8 W/m² yields the correct mean surface temperature and planetary albedo (288 K, 0.3) for the Earth. I further parameterized this for higher and lower temperatures by applying equation (3), which provides a maximum *addir* value slightly higher than Earth's (10 W/m²) at temperatures exceeding ~300 K while assuming that the influence of water clouds on thermal energy balance is negligible at latitudes for which temperatures decrease below the freezing point of water.

$$addir = \max\left(-6.8\log\left(F_C/F_E + 1\right), -10\right) \qquad (3)$$

With equation 3, I am effectively assuming a weak negative cloud feedback in the model for all warm surface temperatures (it is zero otherwise). However, the sign of cloud feedback for conditions removed from Earth's is highly uncertain, with some models finding a positive feedback close to the runaway whereas others finding a negative one (e.g., Leconte et al. 2013; Wolf and Toon 2014). Sensitivity studies setting *addir* to zero had a very small effect on inner edge boundaries. Thus, I see eqn. 3 as a somewhat uncertain, but reasonable conservative estimate.

When temperatures become cold enough for $CO_2$ clouds to form, as they do near the outer edge, I assume a constant cloud fraction of 50%, following GCM studies (Forget et al. 2013).

For both types of clouds, cloud albedo is assumed to be a linear function of zenith angle (Williams and Kasting 1997; Vladilo et al. 2013)(equation 4):

$$a_c = \alpha + \beta z \qquad (4)$$

Here, $a_c$ is cloud albedo, whereas the fitting constants α and β are -0.078 and 0.65, respectively.

Fresnel reflectance data for ocean reflectance at different zenith angles is used. Ice absorption is treated in UV/VIS and near-infrared channels and the proper contribution is calculated for each star type.

I have implemented the following albedo parameterization for snow/ice mixtures, similar to that in Curry et al. (2001) (equation 5)

$$\alpha(visible) = \begin{cases} 0.7 & T \leq 263.15 \\ 0.7 - 0.020(T - 263.15); & 263.15 < T < 273.15 \\ 0.22 & T \geq 273.15 \end{cases}$$

$$\alpha(nir) = \begin{cases} 0.5 & T \leq 263.15 \\ 0.5 - 0.028(T - 263.15); & 263.15 < T < 273.15 \\ 0.22 & T \geq 273.15 \end{cases} \quad (5)$$

The EBM also models water ice coverage within a latitude band ($fice$) to temperature based on empirical data (Thompson and Barron 1981). I find the following fit (equation 6):

$$fice = \begin{cases} 1. & , T \leq 239 \\ 1 - \exp((T - 273.15)/12.5); & 239 < T < 273.15 \\ 0 & , T \geq 273.15 \end{cases} \quad (6)$$

Zonally-averaged, surface albedo at each latitude band is calculated via the following (equation 7):

$$a_s = (1 - f_c)\{(1 - f_o)a_l + f_o[f_i a_i + (1 - f_i)a_o]\} + f_c a_c \quad (7)$$

Here, $a_s$, $a_c$, $a_o$, $a_i$, and $a_l$ are the surface, cloud, ocean, ice, and land albedo, respectively. Likewise, $f_c$, $f_o$, and $f_i$ are the cloud, ocean, and ice fraction, respectively. Following Fairén et al. (2012), at sub-freezing temperatures, the maximum value between ice and cloud albedo is chosen to prevent clouds from artificially darkening a bright ice-covered surface.

Heat transfer efficiency is determined by the following parameterization (equation 8):

$$D = D_o \left(\frac{p}{p_o}\right)\left(\frac{c_p}{c_{p,o}}\right)\left(\frac{m_o}{m}\right)^2\left(\frac{\Omega_o}{\Omega}\right)^2 \quad (8)$$

Here, values with the "$o$" subscript indicate Earth's values. $c_p$ is the heat capacity, $m$ is atmospheric molecular mass, $\Omega$ is rotation rate, $D$ is the globally-averaged heating efficiency, with a value of $D_o = 0.58$ Wm$^{-2}$K$^{-1}$ for the Earth (Williams and Kasting 1997; Ramirez and Levi 2018).

For the outer edge calculations, I found the highest possible CO$_2$ pressure able to maintain a mean surface temperature of 273 K at the farther possible distance from the star. With the inner edge calculations, the CO$_2$ partial pressure was 330 ppm, as per the radiative-convective climate modeling simulations. An aquaplanet with a 23.5 degree obliquity was assumed.

# 3. RESULTS

*Habitable zones: dependence on nitrogen pressure*

I first calculate the HZ limits for 1 bar, 2 bar and 5 bar $N_2$ atmospheres using the radiative-convective climate model (Figure 1). I then compute the same limits using the advanced energy balance model for two different maximum $H_2O$ cloud cover levels (55%, 100%) (Figures 2 – 3). All 3 plots also contain empirical HZ boundaries, the Recent Venus and early Mars limits, respectively, for the inner and outer edge. The Recent Venus limit is based on the observation that water did not exist on the Venusian surface for at least the last billion years (Kasting et al. 1993). In contrast, the early Mars limit is based on geologic evidence suggesting that Mars was a habitable planet ~3.8 billion years ago (Kasting et al. 1993).

As classically performed, the inner edge in the radiative-convective climate model is determined by moving the planet closer to the star until the moist greenhouse is triggered (above ~340 K in a 1 bar $N_2$ atmosphere; Kasting et al., 1993). For the 2-bar and 5-bar $N_2$ atmospheres, the radiative-convective climate model predicts that the moist greenhouse occurs at slightly higher temperatures of ~360 K and 390 K, respectively. This is because additional $N_2$ dilutes the atmosphere, which increases the planetary albedo. For a mean surface temperature of 340 K (for instance), my model predicts a planetary albedo of ~0.2 for a 1 bar $N_2$ background atmosphere but this rises to nearly 0.3 in the 5-bar $N_2$ atmosphere, offsetting a relatively small decrease in outgoing longwave radiation. So, a higher surface temperature would be required to achieve a similarly moist stratosphere for the high $N_2$ pressure planet. This is also why even higher $N_2$ pressures were not considered. The maximum temperature for life on Earth is ~394 K (Clarke 2014).

The 1-bar and 2-bar $N_2$ cases both produce inner edge boundaries that just cross Earth's orbit (Figure 1), consistent with earlier studies (Kopparapu et al. 2013; Kopparapu et al. 2014). This result is obviously incorrect or else Earth would be uninhabitable. As mentioned above, this results from the radiative-convective climate model not including the cooling effects from both sub-saturated relative humidity and clouds.

The higher Rayleigh scattering and lower NIR absorption in planets orbiting hotter stars cause a great variation in the calculated inner edge as a function of $N_2$ pressure. In contrast, the lower Rayleigh scattering and increase in NIR absorption (more efficient atmospheric heating) for planets orbiting cooler stars decrease these variations. At the outer edge, competing cooling and warming effects produce little variation in the computed HZ limits as a function of $N_2$ pressure (Fig. 1), which is consistent with previous work (Kopparapu et al. 2013, 2014) (Figure 1).

Although the computed EBM outer edge limits are very similar to those computed by the radiative-convective climate model, the EBM inner edge limits are located significantly closer to the star (Figures 2 – 3). None of the EBM inner edges cross Earth's orbit (Figures 2 – 3). For a 1 bar $N_2$ atmosphere, the inner edge in our solar system is at ~0.98 and 0.92 AU, respectively, for 55 and 100% maximum cloud cover. This compares well with the 0.93 and 0.95 AU computed by more complex 3-D models (Leconte et al. 2013; Wolf and Toon 2015). For comparison, the 5-bar $N_2$ atmosphere's solar system inner edge is located at ~0.96 and 0.9 AU, respectively, for 55 and 100% maximum cloud cover (Figures 2 – 3).

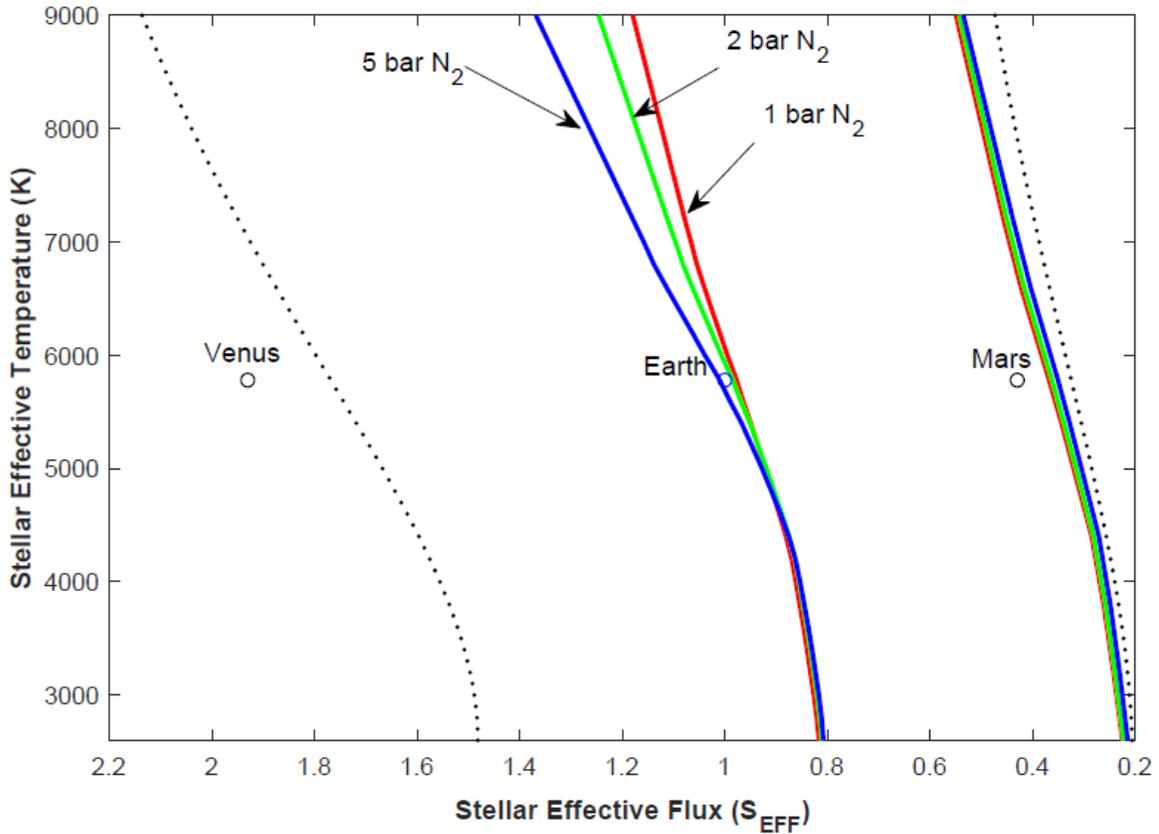

**Figure 1:** Habitable zone limits for A – M stars ($T_{EFF}$ = 2,600 – 9,000K) assuming 1 bar (red), 2 bar (green), and 5 bar (blue) $N_2$ atmospheres using the radiative convective climate model. The empirical "recent Venus" inner and outer edge "early Mars" habitable zone limits (black dotted) are also shown for comparison. Venus, Earth, and Mars are also added.

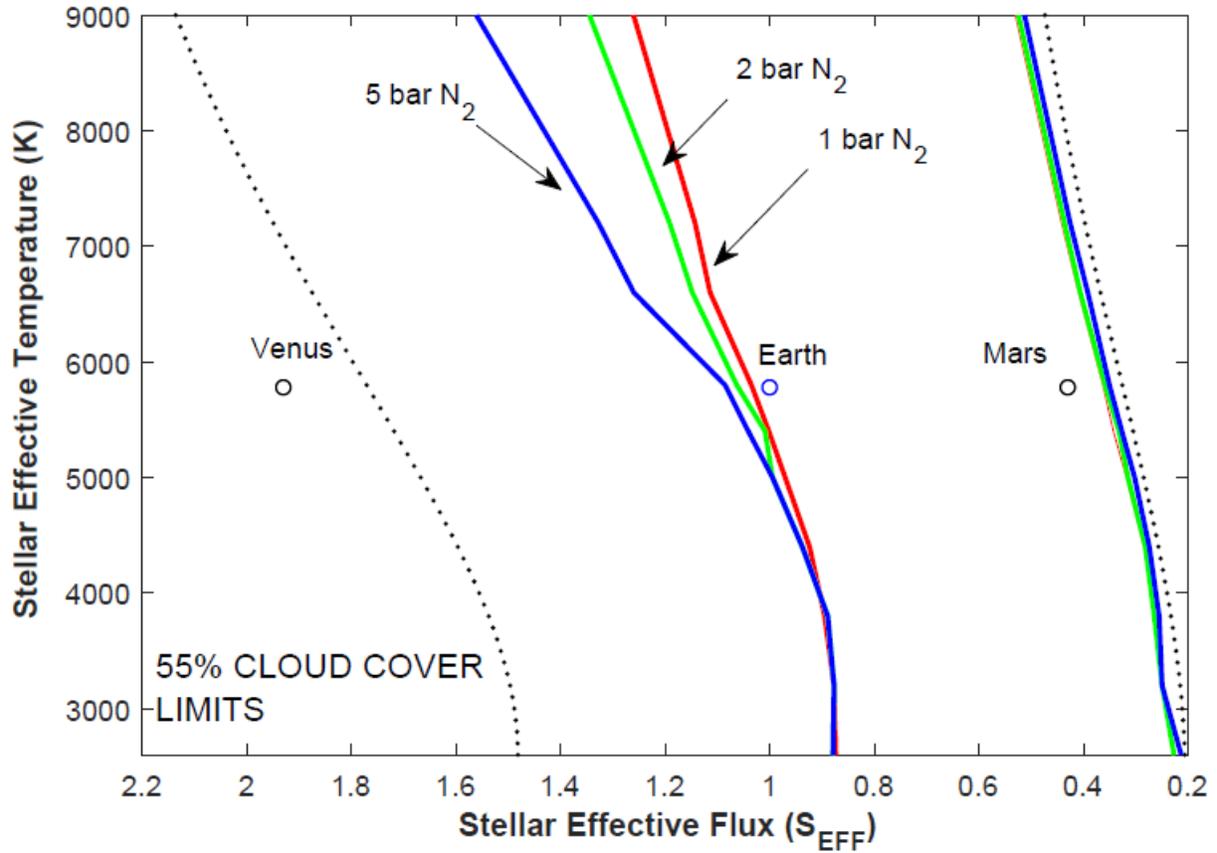

**Figure 2:** Habitable zone limits for A – M stars (T$_{EFF}$ = 2,600 – 9,000K) assuming 1 bar (red), 2 bar (green), and 5 bar (blue) N$_2$ atmospheres using the energy balance model assuming a maximum H$_2$O cloud cover of 55% at the inner edge. The empirical "recent Venus" inner and outer edge "early Mars" habitable zone limits (black dotted) are also shown for comparison. Venus, Earth, and Mars are also added.

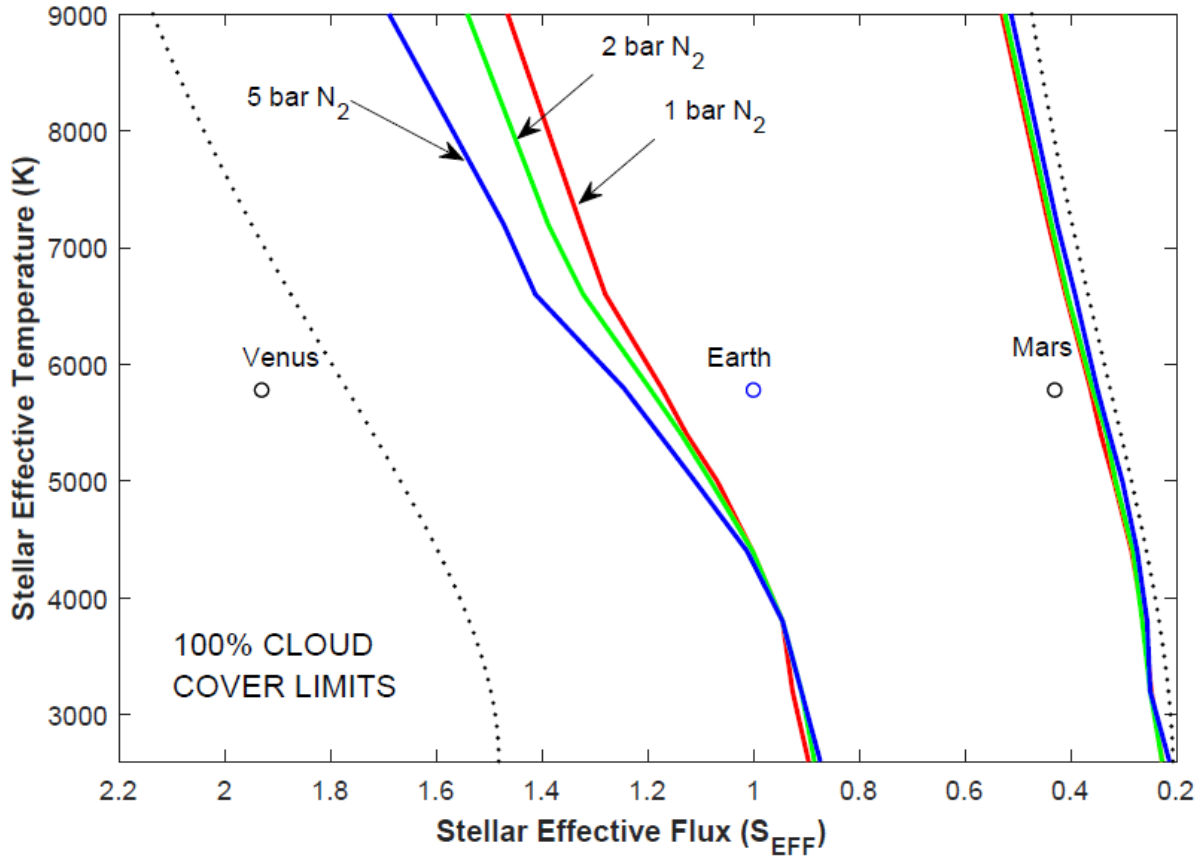

**Figure 3:** Habitable zone limits for A – M stars ($T_{EFF}$ = 2,600 – 9,000K) assuming 1 bar (red), 2 bar (green), and 5 bar (blue) $N_2$ atmospheres using the energy balance model assuming a maximum $H_2O$ cloud cover of 100% at the inner edge. The empirical "recent Venus" inner and outer edge "early Mars" habitable zone limits (black dotted) are also shown for comparison. Venus, Earth, and Mars are also added.

*The nature of the moist and runaway greenhouse*

I also perform EBM calculations to test the existence of the moist greenhouse in 1 bar $N_2$ atmospheres orbiting A-, M-, and Sun-like stars (Figure 4). Instead of triggering a moist greenhouse, both the solar and M-star cases transition into the runaway greenhouse instead. For the runaway greenhouse, the net incoming flux exceeds the longwave thermal flux above a given temperature. In contrast, a proper moist greenhouse is only achieved for the A-star planet. That is, a mean surface temperature of ~ 340 K is achieved when the stratospheric water vapour mixing exceeds ~$3 \times 10^{-3}$ in my model. Typical mean stratospheric temperatures for such conditions range between ~190 - 200 K, similar to the 200 K assumed in radiative-convective climate modeling calculations (see Methods). This is followed by the runaway greenhouse once mean surface temperatures exceed ~360 K for this case (Figure 4). The direct transition to a runaway greenhouse for the planet orbiting our Sun occurs above a mean surface temperature of

~330 K, an almost identical result to that achieved by one 3-D model (Leconte et al. 2013). The runaway greenhouse is triggered above a slightly lower mean surface temperature (~320 K) for the M-star planet. These results are very different from those of radiative-convective climate modeling simulations which predict a moist greenhouse in all cases (Kasting et al. 1993; Kopparapu et al. 2013). These also differ from those of some 3-D models that obtain a moist greenhouse for planets orbiting Sun-like stars (Wolf and Toon 2015; Popp et al. 2016) and M-stars (Kopparapu et al. 2017).

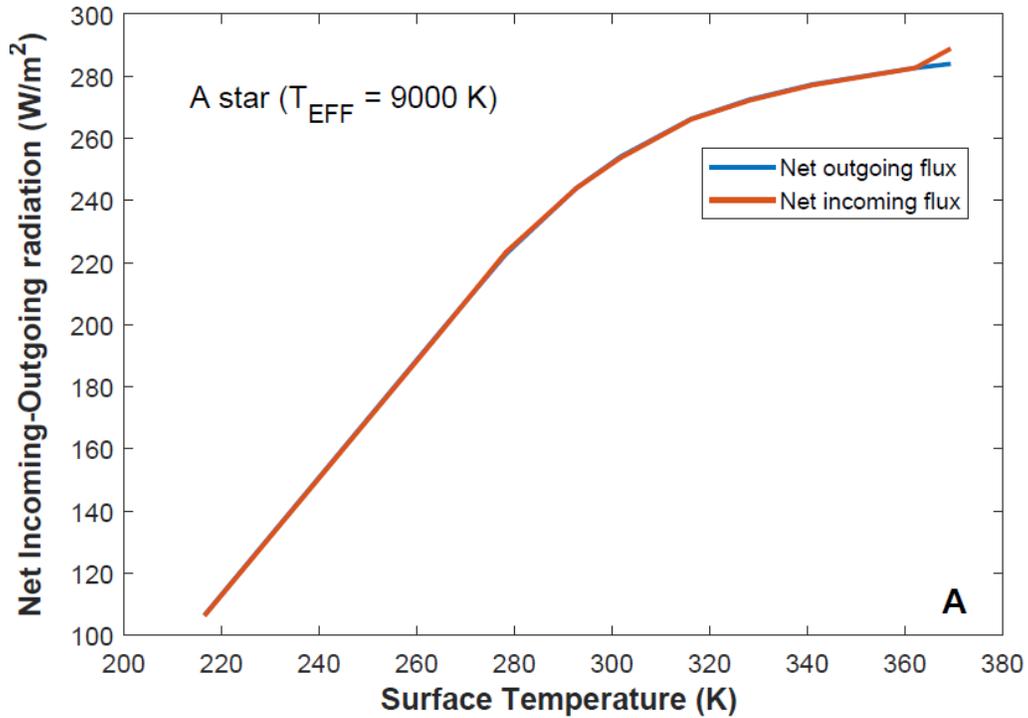

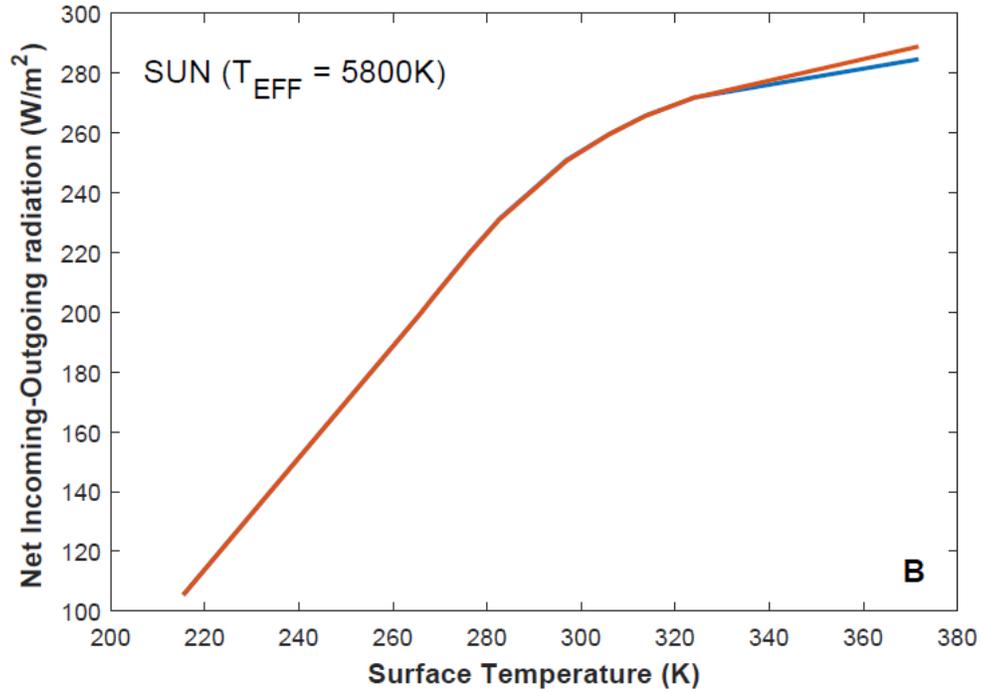

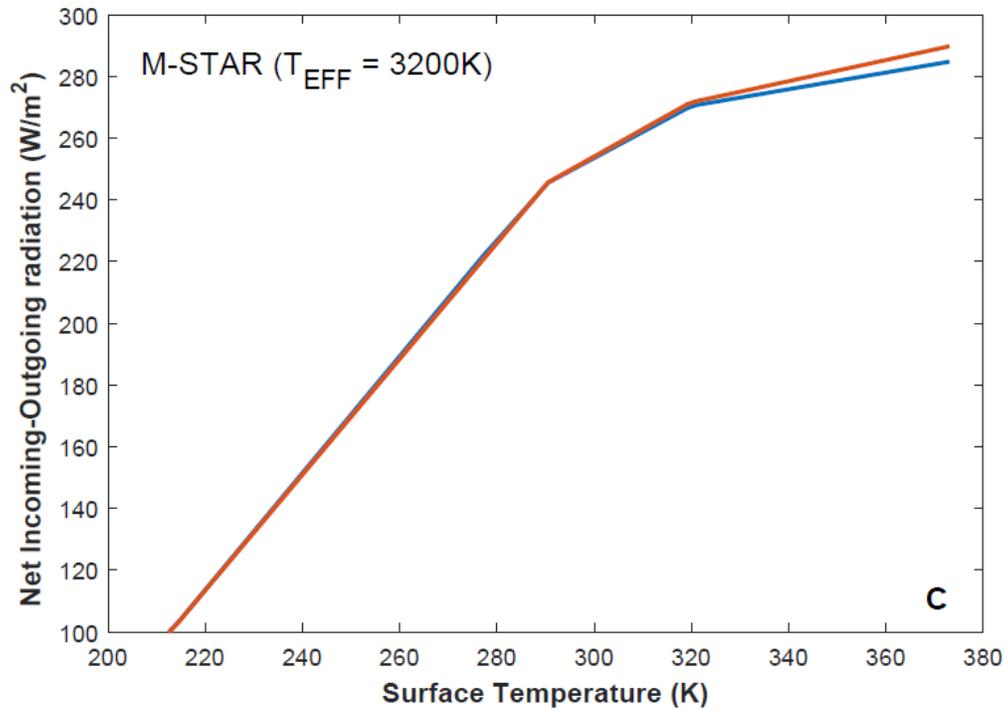

**Figure 4:** Net outgoing (blue) and net incoming stellar (red) fluxes as a planet with a 1 bar $N_2$ atmosphere is moved toward the inner edge of an (a) A star, (b) Sun, and (c) M-star. The moist greenhouse is only triggered for the A-star planet, which occurs in this model when mean surface temperature exceeds ~340 K. The asymptotic outgoing longwave flux is ~284 W/m$^2$, a similar value to that computed in other models (Goldblatt et al. 2013; Leconte et al. 2013). A moist greenhouse for planets orbiting F-stars (not shown) was not found either.

Recent calculations suggest that the moist greenhouse is triggered at lower surface temperatures ($<$ ~300 K) on synchronously-rotating planets (Fujii et al. 2017; Kopparapu et al. 2017), particularly those orbiting M-dwarfs with high near-infrared emission (Fujii et al. 2017). Once near-infrared planetary absorption is high enough, stratospheric temperatures increase as well, which can trigger a moist greenhouse at lower temperatures (Ramirez 2018b). I test this idea with the EBM by increasing calculated stratospheric temperatures by 30% (which gave mean global stratospheric temperatures of ~ 250 K at the destabilization threshold) and repeating the above moist/runaway greenhouse calculation for the solar and M-star cases (Figure 5). In this computation, I only demonstrate the case in which maximum cloud cover is 55% because the 100% cloud cover case exhibits similar trends.

As with the baseline calculation above, a runaway greenhouse is triggered whereas the moist greenhouse is not (Figure 5). Again, this result differs from that in radiative-convective climate modeling calculations (Kasting et al. 1993; Kopparapu et al. 2013; Ramirez 2018b). Nevertheless, in agreement with recent 3-D models (Fujii et al. 2017; Kopparapu et al. 2017), the destabilizing threshold does occur at lower mean surface temperatures, ~290 K and 320 K for the M-star and Sun cases, respectively (Figure 5), only that the destabilization is a runaway greenhouse instead. The other interesting result is that the runaway greenhouse occurs at distances closer to the star for the high stratospheric temperature scenarios. The runaway greenhouse $S_{EFF}$ for the baseline 1 bar $N_2$ M-star case increases from ~0.87 to 0.99 (0.112 to 0.105 AU, assuming L = 0.011$L_{sun}$). For the Sun, these values increase from ~1.03 to 1.18 (0.98 to 0.92 AU). This is because of greater longwave emission at higher stratospheric temperatures, which supports stable climate states at a higher $S_{EFF}$ (Figure 5). For instance, at 300 K, the high stratospheric temperature planet orbiting the Sun-like star has an outgoing longwave emission of ~290 W/m$^2$ as opposed to only ~255 W/m$^2$ for the standard case (Figure 4b; 5b). Thus, unlike previous studies on synchronously rotating planets (Fujii et al. 2017; Kopparapu et al. 2017), the current model predicts that this cooler destabilization threshold can also occur in rapidly-rotating planets.

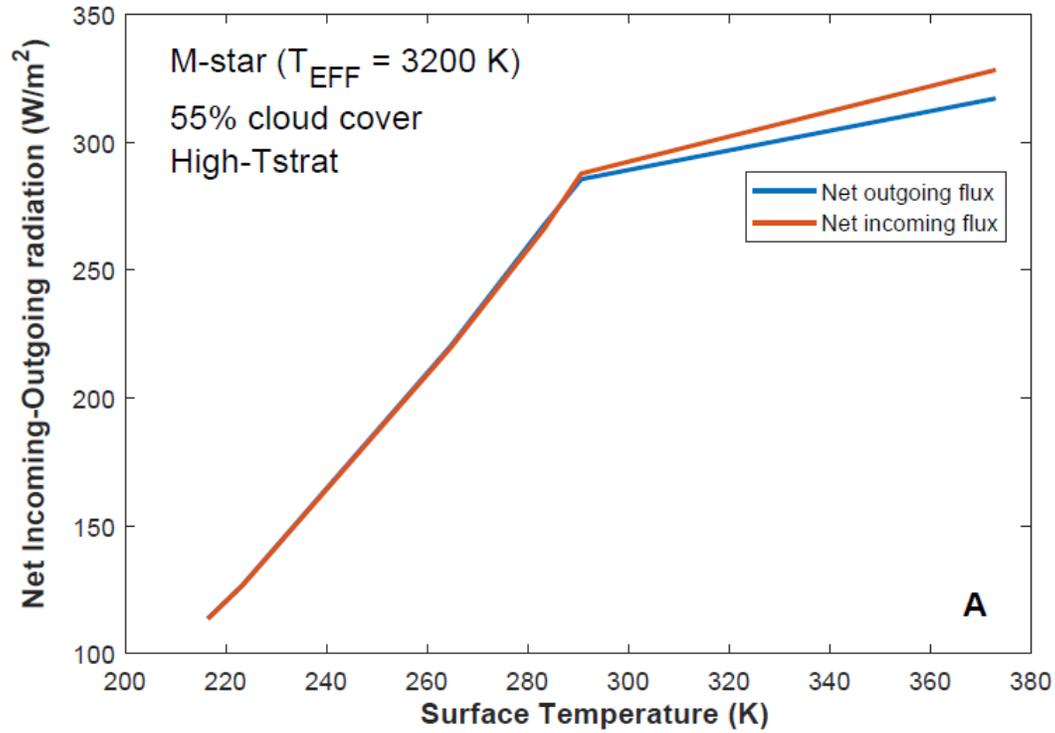

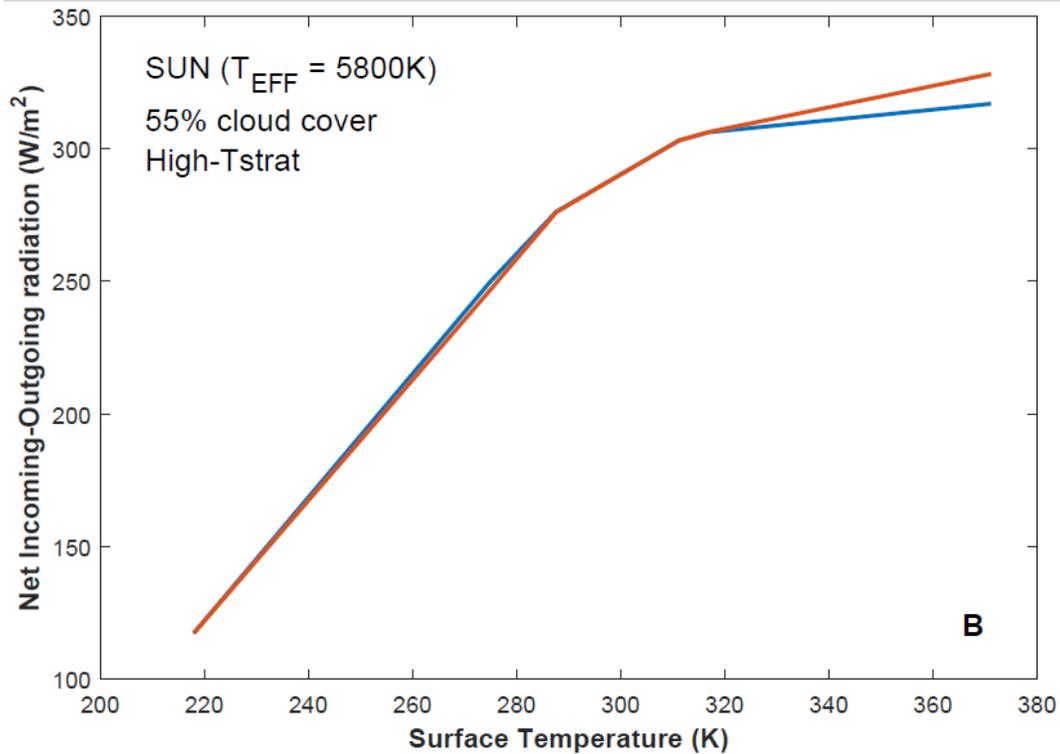

**Figure 5:** Net outgoing (blue) and net incoming stellar (red) fluxes for a planet under the high temperature stratospheric scenario (see text) and 1 bar $N_2$ atmosphere is moved toward the inner edge of an (a) M star and (b) Sun. The resultant increased emission and higher planetary albedo produce an increase in the asymptotic outgoing longwave radiation (> 300 W/m²).

For the Sun, I also assessed other ways in which the inner edge can be moved closer to the star than in the baseline calculation. At 100% cloud cover, the $S_{EFF}$ for the high stratospheric temperature scenario increased further from 1.18 to 1.35 (~0.92 to 0.86 AU). At this closer distance to the Sun, the runaway greenhouse was triggered above a low ~302 K (Figure 6a), lower than the ~320K for the 55% cloud cover high stratospheric scenario (Figure 5b). For the baseline stratospheric scenario, the runaway greenhouse in the 5 bar $N_2$ background pressure case was triggered above ~295 K (Figure 6b). Thus, both a high planetary albedo (from higher cloud cover) or high stratospheric temperatures (higher emission) can reduce the climate destabilization temperature.

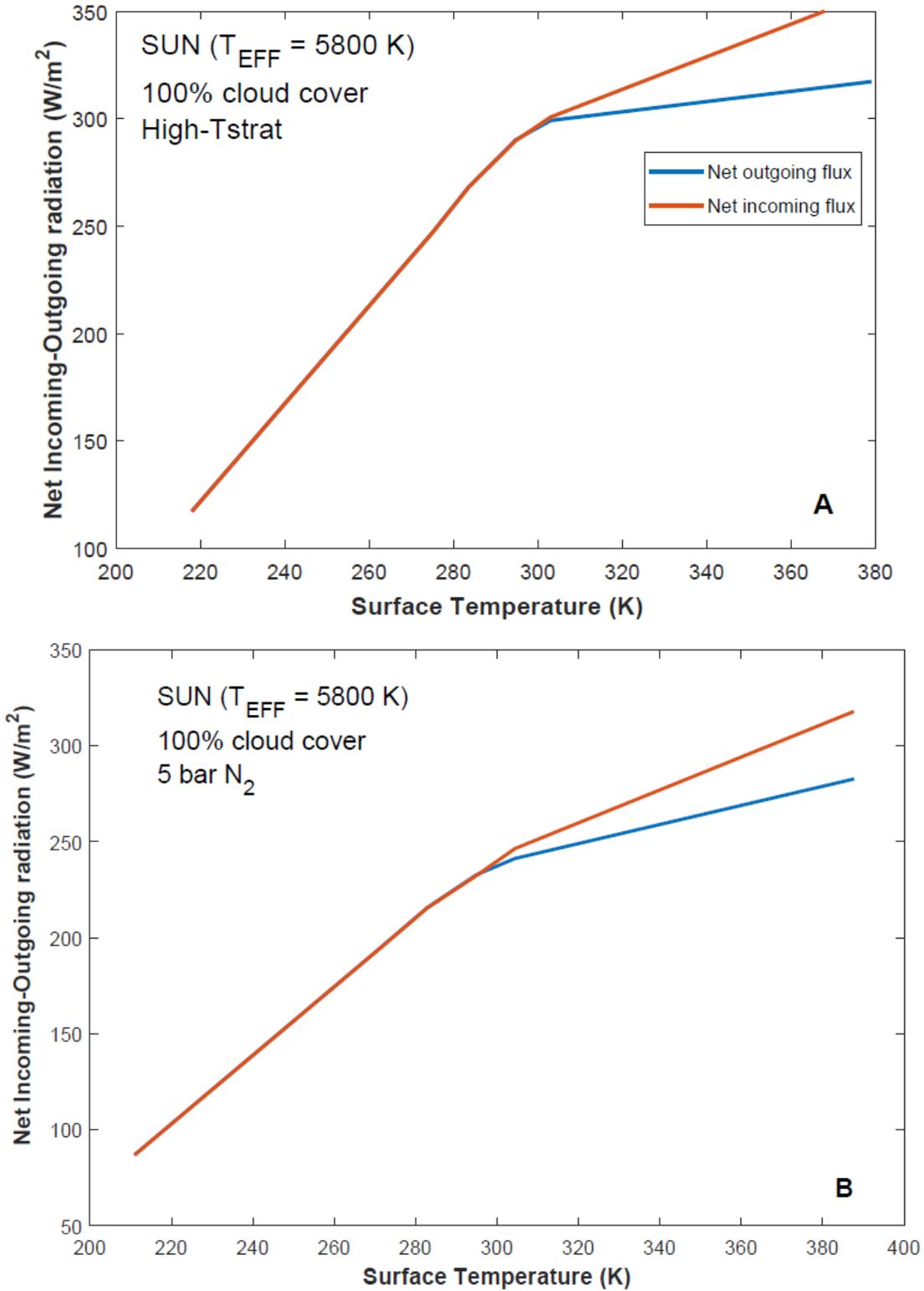

**Figure 6:** Net outgoing (blue) and net incoming stellar (red) fluxes as a planet is moved toward the inner edge of our solar system assuming a (a) 1 bar $N_2$ high stratospheric temperature scenario (explained in text) and (b) a 5 bar $N_2$ atmosphere, respectively. A 100% maximum cloud cover is assumed at the inner edge.

A visualization of how planetary albedo evolves as a planet is pushed closer to the inner edge as a function of maximum cloud cover or background nitrogen pressure is shown in Figure 7. In all cases, the planetary albedo starts high far away, surfaces are icy, and mean surface temperatures are low. The subsequent increase in water vapour absorption and surface temperature reduces the planetary albedo until this is reversed by high Rayleigh scattering in an optically thick atmosphere. This increase in planetary albedo from either high cloud cover or high atmospheric $N_2$ maintains surface temperatures clement, even at relatively small orbital distances. In contrast, relatively weak cooling mechanisms keep planetary albedo relatively low in the 1 bar $N_2$ 55 per cent cloud cover case (Fig. 7), making the runaway greenhouse possible out to relatively larger orbital distances.

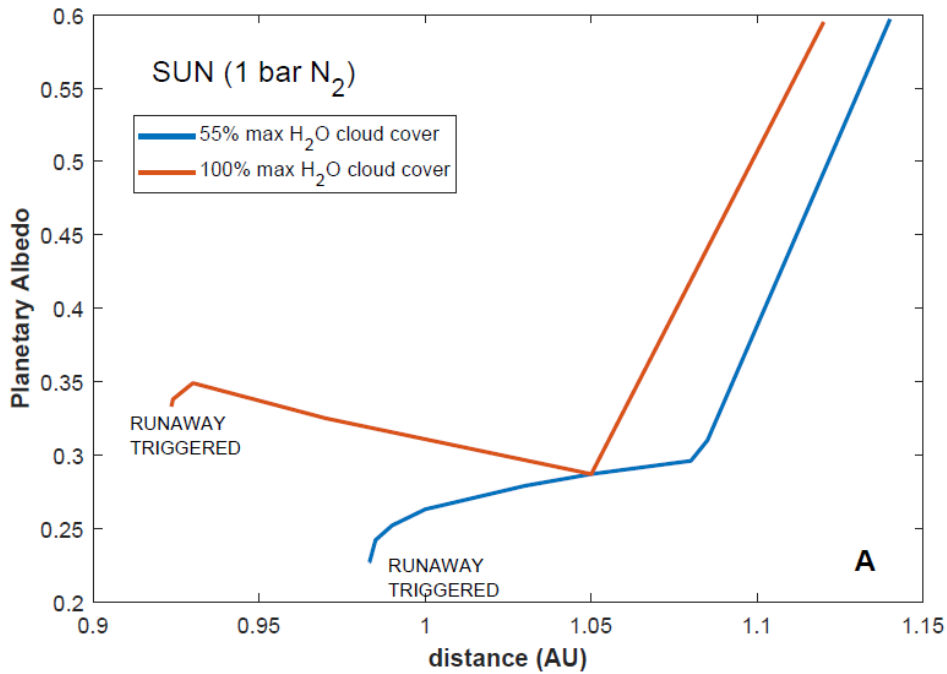

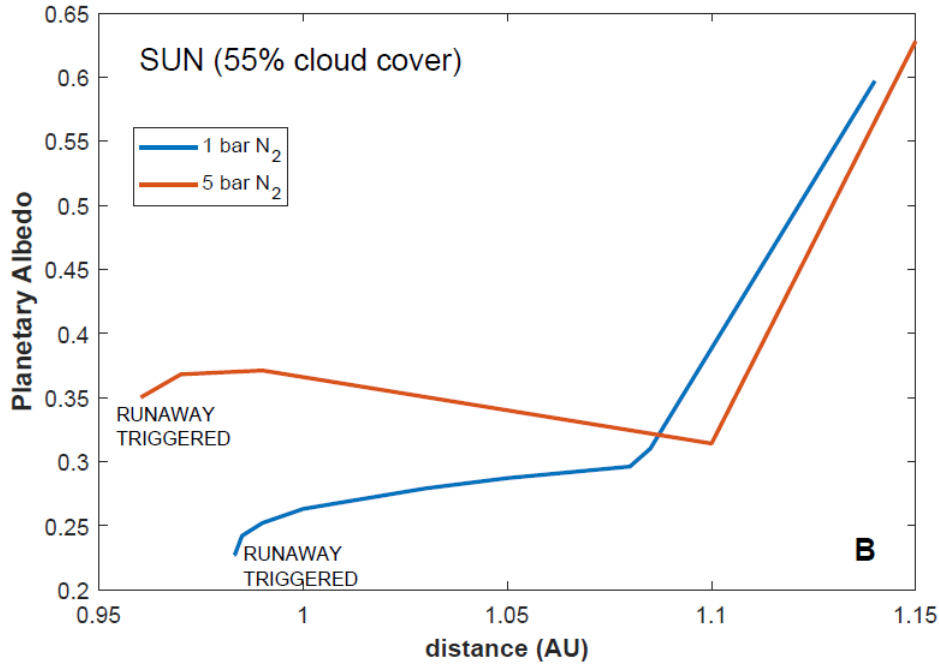

**Figure 7:** Planetary albedo versus semi-major axis distance for a a) 1 bar $N_2$ atmosphere planet orbiting the Sun that achieves a maximum cloud cover of (blue) 55% and (red) 100%, respectively and for a (b) planet that achieves a maximum cloud cover of 55% but with background $N_2$ pressures of (blue) 1 bar and (red) 5 bar, respectively.

## 4. DISCUSSION

*4.1 Comparison between 1-D and 3-D models on the habitable zone boundaries and moist greenhouse threshold*

The results here suggest that outer edge distances as computed by radiative-convective climate models may be robust, even at high nitrogen pressures. Three dimensional models obtain outer edge results that are also consistent with those from 1-D models (Wolf 2018). The situation is very different for inner edge calculations, however, as differences in convective, cloud, and radiation schemes produce a wide scatter across different 1-D and 3-D models (Yang et al. 2013; Kopparapu et al. 2016; Kopparapu et al. 2017; Bin et al. 2018; Ramirez 2018a). These differences partially arise from the difficulty in simulating the water vapour feedback at higher temperatures, which is not a major concern at the HZ outer edge, where water vapour absorption is weaker (Godolt et al. 2016; Ramirez and Levi 2018). Nevertheless, the trend that higher background $N_2$ pressures push the inner edge closer to the star would likely be found in many other models as well.

Overall, the EBM results are consistent with a previous 3-D model suggesting that the moist greenhouse state may not be achievable in our solar system (Leconte et al. 2013). I also find that a moist greenhouse may not be possible on planets orbiting most star types, except for those around very hot stars (e.g., A-stars). The moist greenhouse on HZ planets orbiting extremely hot stars can be linked to reduced NIR absorption and less efficient stellar absorption (Fig. 4.), preventing the runaway from occurring at lower surface temperatures. Such moist and runaway greenhouse thresholds cannot be reliably determined using the inverse method (see Methods) in radiative-convective climate models because that technique ignores energy balance and only evaluates the relative change in the energy budget as the planet moves closer to or farther away from the star. However, radiative-convective climate modeling simulations that explicitly achieve radiative-convective equilibrium, and with a proper relative humidity/cloud parameterization, could perform a similar calculation (e.g., Ramirez et al. 2014a).

*4.2 The presence of global surface temperature inversions in very warm convective atmospheres*

These results, and those of Leconte et al. (2013), seemingly contradict those found in Wolf and Toon (2015). In that study, a moist greenhouse for an Earth analogue in our solar system is found, although that model, which is an earlier version of subsequent models (e.g., Kopparapu et al. 2017), exhibits additional behaviour not seen in this or the Leconte et al. (2013) study. Wolf and Toon (2015) obtain a growing mean surface temperature inversion for global mean surface temperatures ~330 K and above, roughly the same mean surface temperature that this model and that of Leconte et al. (2013) compute for the runaway greenhouse instead (a smaller inversion is also seen in Popp et al., 2016).

Although surface inversions do occur on Earth, these are temporary (e.g. night-time) or occur on a local or regional scale, as with polar inversions during the cold months (Hartmann 1994). For polar inversions, latent heat fluxes are essentially zero because of very low surface temperatures. Nevertheless, both types of inversions are sustained by additional energy from other locations (e.g. meridional, lateral heat transport). In effect, Earth does not exhibit surface temperature inversions on a global, annually-averaged scale, even should they occur regionally and at specific times. In contrast, the Wolf and Toon (2015) surface temperature inversions are permanent and large enough to encompass much, if not nearly all of, the planet.

Additional insights can be gleaned from surface energy balance, which states that the latent ($F_L$) plus sensible ($F_S$) heat fluxes are balanced by the net thermal radiation (e.g. convective heat flux)(e.g., Ramirez et al. 2014a) (eqns. 9 – 10). I will ignore the ground heat storage term for simplicity.

$$F_L + F_s = F_{IR} + F_{stellar} \qquad (9)$$

$$F_L + c_p \rho_{atm} C_D v (T_s - T_{atm}) = \sigma (T_{atm}^4 - T_s^4) + F_{stellar} \qquad (10)$$

Here, $c_p$ is the atmospheric heat capacity, $C_D$ is a drag coefficient, $v$ is the near-surface wind velocity, $T_s$ is the surface temperature, $T_{atm}$ is the near-surface atmospheric temperature, $F_{IR}$ is the outgoing longwave flux, and $F_{stellar}$ is the net absorbed stellar flux.

Wordsworth and Pierrehumbert (2013) defend the inversion by arguing that $T_{atm} > T_s$, which requires that the latent heat flux ($F_L$) be positive for eqn. (10) to exactly balance, triggering surface evaporation (and therefore, convection). In that case, the sensible heat flux term is negative.

However, in contrast to Wordsworth and Pierrehumbert (2013), Wolf and Toon (2015) do not generally obtain convection in the inversion region (although they obtain some limited convection in their cooler moist greenhouse simulations) . This follows because convection cannot proceed if the air and the ground are both hotter than the near-surface air in between (Kasting et al. 2015). The ground is warmer than the near-surface air so $T_s$ *must* be greater than $T_{atm}$ in eqn. (10).

In the Wolf and Toon (2015) scenario, the moist greenhouse atmosphere gets so optically thick that the convective energy flux ($F_L + F_s$) approaches 0 and the inversion layer becomes opaque to radiation (see eqns. 9 - 10), which is again consistent with no near-surface convection. They argue that convection would instead occur in the upper atmosphere, well above the inversion layer. However, it remains unclear how the upper atmosphere becomes wet enough to trigger and sustain a moist greenhouse in the absence of water vapour evaporating from a large surface reservoir (i.e. oceans).

To demonstrate, the following equation is typically used to estimate the escape rate in moist greenhouse atmospheres, assuming escape is diffusion-limited (e.g., Kasting et al. 1993):

$$\phi = \frac{b \cdot f_H}{H} \sim 2.3 \times 10^{13} \cdot f_H \text{ particles/cm}^2/\text{sec} \tag{11}$$

Where, $b$ is a diffusion parameter, $H$ is the homopause scale height, and $f_H$ is the hydrogen mixing ratio at the homopause. At the homopause, $f_H$ is approximately twice the mixing ratio of water vapour (~$6 \times 10^{-3}$ for the moist greenhouse). Earth's oceans contain ~$2 \times 10^{28}$ H particles/cm$^2$, so an entire Earth ocean can escape to space within ~ 4.5 Gyr (the age of the solar system) according to equation (11). However, if the lower atmosphere is cut off from the rest of the atmosphere, then the available water that can escape to space is severely diminished. Wolf and Toon (2015) suggest that a typical moist greenhouse atmosphere in their model may have a global water vapour column of ~500 kg/m$^2$, which translates to ~ $1.8 \times 10^{-4}$ Earth oceans worth of water vapour. If we assume half of that is above the inversion layer, the upper atmosphere would be completely desiccated in well under 500,000 years (~0.01% the age of the solar system), ending the moist greenhouse episode, unless the lower atmospheric circulation is able to eventually reconnect to the upper atmosphere. Thus, if the Wolf and Toon (2015) scenario is correct, moist greenhouses may exist on some planets around Sun-like stars, but in a very fleeting form, requiring great serendipity for astronomers to detect their activity from Earth.

Moreover, there is evidence that the presence or absence of such surface inversions could explain why a moist greenhouse is found in some models but not in others. The planetary albedo for an Earth analogue in Wolf and Toon (2015) decreases from ~0.35 at a present day solar flux to

~0.26 at a mean surface temperature of ~330 K. After this, the planetary albedo starts to increase, coinciding with the appearance and subsequent growth of the surface inversion. Planetary albedo reaches a maximum value of ~0.33 at a mean surface temperature of ~363 K. This increase in reflectivity is sufficient to delay the runaway greenhouse, allowing the moist greenhouse to be triggered first. In contrast, the planetary albedo for our 1 bar $N_2$ Earth analogue becomes significantly lower (~0.23), triggering a runaway greenhouse at a lower mean surface temperature (~330 K) (Figure 4b; Figure 7), as in the Leconte et al. (2013) model. Therefore, without the cooling afforded by the surface inversion, the planetary albedo in the Wolf and Toon (2015) model would have likely continued to decrease slightly until the net absorbed solar flux exceeded the threshold, triggering the runaway greenhouse in place of a moist greenhouse.

In summary, future work should evaluate whether permanent nearly-global temperature inversions in moist greenhouse atmospheres are plausible or if they represent a breakdown in the model physics. The presence of such inversions may also help explain why the moist greenhouse threshold is predicted in some models while a runaway greenhouse is found in others.

*4.3 Low temperature threshold for moist and runaway greenhouses*

The results here also confirm the idea that higher stratospheric temperatures can dramatically decrease the runaway or moist greenhouse threshold temperature for planets (< ~300 K) (Ramirez 2018b). These results are also consistent with those in Fujii et al. (2017), whom attribute the increase in stratospheric temperatures to enhanced NIR absorption on M-star planets. That said, the current model cannot test how such high stratospheric temperature scenarios may arise. More studies with appropriate convective schemes should analyze the moist and runaway greenhouse regime for M-stars.

Nevertheless, my model finds that other mechanisms, including reflective clouds or haze (unless the cloud feedback at higher temperatures is overly positive) and high background $N_2$ pressures, can also reduce the runaway greenhouse temperature threshold. In these mechanisms, it is the higher planetary albedo that allows planets to be moved closer to their stars. For the high stratospheric temperature cases, it is increased longwave emission that permits stable climate states at a higher $S_{EFF}$. In all cases, once the ability of the planet to cool itself is exceeded (e.g., the cloud coverage cannot increase beyond 100%), it immediately transitions into the runaway greenhouse (Figures 5 – 7). At such close distances to the star, the $S_{EFF}$ value is so high that rapid destabilization is triggered from a relatively low mean surface temperature. Moreover, such mechanisms can occur on planets orbiting any star type. The implication then is that the inner edge for any star type may be closer to the star than previously thought.

*4.4 Other solar system and extrasolar system implications*

The HZ plots also suggest that the inner edge may be easiest to define for M-dwarfs as the low and high pressure $N_2$ cases converge to similar $S_{EFF}$ values for planets orbiting such stars (Figures 1 – 3). In contrast, relatively high Rayleigh scattering and low NIR absorption for planets orbiting hotter stars cause more scatter between the high and low $N_2$ pressure cases. That said, clouds remain a major influence for all star types as shown here and in previous work (Yang et al. 2013). For instance, for tidally-locked M-star planets, 3-D models predict that a highly-reflective cloud deck forms on the substellar point (Yang et al. 2013; Kopparapu et al. 2017; Bin et al. 2018), which may push the planet as close or closer to the star that what is predicted here for rotating planets (depending on negative cloud feedback strength). This cooling mechanism, allowing destabilization to occur at a higher $S_{EFF}$ (closer to the star), may also contribute to the lower temperature moist greenhouse threshold for M-dwarfs HZ planets observed in some works (Fujii et al. 2017; Kopparapu et al. 2017).

The results here also have interesting implications for the evolution of Venus. Assuming Venus had formed with a significant water inventory, it could have lost it in a moist or runaway greenhouse (Kasting 1988). Some recent work support the latter possibility, likely early in its history (e.g., Hamano et al. 2013; Ramirez and Kaltenegger 2014). The current work not only supports a runaway greenhouse being the ultimate demise of Venus, but the "Recent Venus" limit is also much closer to the star than any of the revised inner edge limits computed here, suggesting that Venus lost its water well before ~1 Gyr ago (assuming the world began with a significant water inventory to begin with). That said, if rotation rates on early Venus were slow enough, desiccation could have been completed somewhat later in history (Way et al. 2016).

*4.5 Model differences and caveats*

Lastly, this analysis assumes a weak negative feedback at higher temperatures (see Methods). This is consistent with Ramirez et al. (2014a), whom have argued that surface relative humidity should increase at higher temperatures for fast-rotating planets, which would increase the amount of low water clouds and enhance this negative cloud feedback. A similar increase in surface relative humidity with surface temperature was also found in Wolf and Toon (2015). However, as the moist greenhouse was triggered, their model's lower atmosphere became drier and relative humidity decreased. In contrast, other models obtain the opposite result, or a positive cloud feedback at higher surface temperatures than the Earth (Leconte et al. 2013). In that model this positive cloud feedback was countered by a negative feedback from a weakening Hadley circulation. Such model discrepancies, which can have a big impact on the existence and nature of moist and runaway greenhouses, may be attributable to differences in water vapour transport or convection schemes. To summarize, both convection schemes and clouds remain poorly-understood for regimes far removed from that of present day Earth's, making continued improvement of associated climate model physics key to further advancement.

**5. CONCLUSION**

An advanced energy balance model (EBM) with clouds is used to derive new habitable zone boundaries for A – M stars for a number of atmospheric $N_2$ pressures (1, 2 and 5 bar). The outer edge boundaries for the three atmospheric pressures are consistent with those predicted by radiative-convective climate modeling simulations. With the diminished water vapour feedback, the outer edge limits change very little as a function of $N_2$ pressure. However, relatively large differences exist at the inner edge because of a stronger water vapour feedback, high planetary albedo, and Rayleigh scattering effects. In the 5 bar $N_2$ atmosphere case, the habitable zone extends from ~0.9 – 1.7 AU in our solar system, which is significantly wider than for the classical 1 bar $N_2$ case (~1 – 1.67 AU).

The EBM also finds that planets around F – M stars all directly transition to a runaway greenhouse and never enter a moist greenhouse. Only for planets orbiting A-stars does a bona-fide moist greenhouse occur preceding the runaway. Thus, these results are consistent with those of 3-D models suggesting that only a runaway greenhouse would be triggered on Earth at higher temperatures (Leconte et al. 2013). In agreement with Leconte et al. (2013), this would occur on our planet above a mean surface temperature of ~330 K.

Consistent with previous results (Fujii et al. 2017; Ramirez 2018b), the model finds that high stratospheric temperatures can reduce the mean surface temperature required to trigger a moist or runaway greenhouse. Higher longwave emission in this case allows potentially habitable planets closer to the star that what would be the case.  However, I also find that other cooling mechanisms as the planet is pushed closer to the star, including highly-reflective clouds or high $N_2$ atmospheric pressures, through an increase in planetary albedo, can also reduce the temperatures required to trigger a runaway greenhouse.  These cooling mechanisms, which can manifest around any star type, also decrease the inner edge distance, widening the habitable zone.

Finally, I argue that the appearance of near-global permanent surface temperature inversions in moist greenhouse atmospheres may be inconsistent with our understanding of atmospheric convection and energy balance. They may also present challenges for the future detection of moist greenhouse activity in such atmospheres. Future work should reassess the plausibility of such scenarios.

## ACKNOWLEDGEMENTS

R.M.R. acknowledges support by the Earth-Life Science Institute (ELSI) and the National Institutes of Natural Sciences: Astrobiology Center (grant number JY310064). The author acknowledges constructive comments by Ravi kumar Kopparapu. R.M.R. also had discussions with several other investigators, including Eric Wolf, Jim Kasting, Yuka Fujii, and Tony Del Genio.